\documentclass[a4paper]{article}

\usepackage{INTERSPEECH2020}
\usepackage{amsmath,graphicx}
\usepackage{latexsym}
\usepackage{CJKutf8}
\usepackage[T1]{fontenc}
\usepackage{multirow}
\usepackage{amssymb}
\usepackage{subfig, graphicx}
\usepackage{url}
\usepackage{booktabs}

\title{Learning Fast Adaptation on Cross-Accented Speech Recognition}
\name{Genta Indra Winata$^{\star\dagger}$, Samuel Cahyawijaya$^{\star}$, Zihan Liu$^{\star}$, Zhaojiang Lin, \\Andrea Madotto, Peng Xu, Pascale Fung\thanks{$^\star$Equal contributions.}}
\address{
  Center for Artificial Intelligence Research (CAiRE)\\
  The Hong Kong University of Science and Technology}
\email{$^\dagger$giwinata@connect.ust.hk}

\begin{document}

\maketitle

\begin{abstract}
Local dialects influence people to pronounce words of the same language differently from each other. The great variability and complex characteristics of accents creates a major challenge for training a robust and accent-agnostic automatic speech recognition (ASR) system. In this paper, we introduce a cross-accented English speech recognition task as a benchmark for measuring the ability of the model to adapt to unseen accents using the existing CommonVoice corpus. We also propose an accent-agnostic approach that extends the model-agnostic meta-learning (MAML) algorithm for fast adaptation to unseen accents. Our approach significantly outperforms joint training in both zero-shot, few-shot, and all-shot in the mixed-region and cross-region settings in terms of word error rate.
\end{abstract}

\section{Introduction}
Spoken languages show great variation across regions and such distinctions derive from the phonetics of local dialects and language backgrounds. Despite the high performance reported by state-of-the-art English automatic speech recognition (ASR) systems, accented speech recognition is still an unsolved real-world challenge due to the great variability of accents and their complex characteristics~\cite{kat1999fast}. It is difficult for ASR models to adapt to unseen accents which have relatively distinct pronunciations and tones compared to the accents used for training the ASR models. Increasing the number of training data and exposing the model to different accents is a common solution to improve the model's robustness to different speakers' accents by introducing variations. However, such approaches are costly and not scalable due to the difficulties in collecting high-quality speech data with different accents. Existing data augmentation techniques such as noise injection~\cite{narayanan2014joint} and speed perturbation~\cite{hori2017advances} have been proposed to overcome the limitation on high-resource data. In this work, we explore training approaches for fast adaptation to unseen accents instead of augmenting the training data.
We apply model-agnostic meta-learning (MAML)~\cite{finn2017model} to teach the model to learn new tasks faster and more efficiently, and our approach can easily be applied to few-shot learning. A few studies have explored joint and multi-task training on multiple accent speech recognition models~\cite{sun2018domain,jain2018improved,jain2019multi}. However, none thoroughly investigated few-shot learning on the cross-accented speech recognition task.




In this paper, we introduce a cross-accented speech recognition task derived from existing dataset, CommonVoice~\cite{ardila2019common}, to move toward building a robust speech recognition system. The motivation of this work is to establish a benchmark for evaluating cross-accented speech recognition. We introduce an accent-agnostic model by applying meta-learning as a learning to learn method for fast accent adaptation. The trained model is able to rapidly adapt to recognize speech with unseen accents. We train our transformer~\cite{vaswani2017attention} speech recognition model on a set of accents via meta-learning and fine-tune the trained model with a few samples of target accented speech. Experimental results show that our approach is able to quickly adapt to new accents more effectively than joint-training, and interestingly, our approach is also able to handle zero-shot predictions. 


\begin{figure}[!t]
\begin{minipage}{.23\textwidth}
  \centering
  \includegraphics[width=1.15\linewidth]{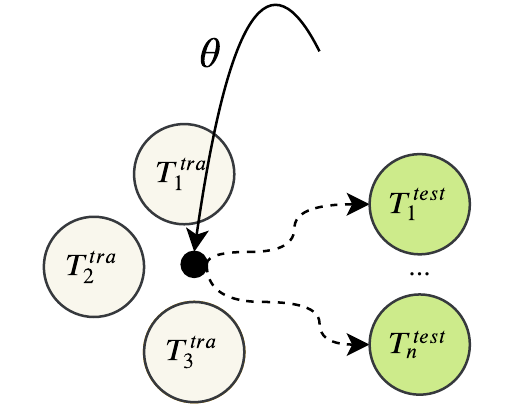}
\end{minipage}
\begin{minipage}{.23\textwidth}
  \centering
  \includegraphics[width=1.02\linewidth]{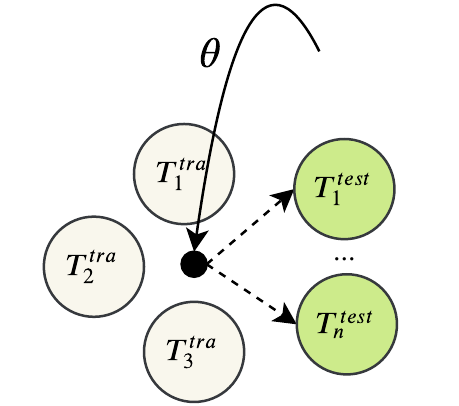}
\end{minipage}
\caption{Illustration modified from~\cite{gu2018meta} of fine-tuning from joint training \textbf{(left)} and meta-learning \textbf{(right)}. The solid line represents the model optimization path of the initial parameters and dashed line represents the fine-tuning path. The white off circles are training accents and green circles are testing accents.}
\label{fig:maml}
\end{figure}

\section{Related Work}

\subsection{Meta-Learning}
Meta-learning is a sub-field of machine learning that designs models to learn new tasks in a new setting with a few training examples
~\cite{schmidhuber1992learning,thrun2012learning}. In a recent work, \cite{finn2017model} propose model-agnostic meta-learning (MAML) and show the application of meta-learning in a deep learning framework. Several meta-learning-based models have been proposed for solving few-shot image classification
\cite{ravi2016optimization,vinyals2016matching,santoro2016meta} and natural language processing applications, such as text classification~\cite{yu2018diverse}, dialogue response generation~\cite{madotto2019paml,qian-yu-2019-domain}, low-resource machine translation~\cite{gu2018meta}, semantic parsing~\cite{huang2018natural}, and sales prediction~\cite{lin2019learning}. \cite{gu2018meta} makes the interesting finding that MAML actually is able to generalize the model in the low-resource machine translation task without any fine-tuning steps or when there is no information on the target accent. In speech applications, \cite{hsu2019meta} introduce the practicality of applying MAML in cross-lingual speech recognition, while in another line of works, MAML has been applied to learn how to adapt respectively to the speaker \cite{klejch2018learning,klejch2019speaker}.

\subsection{Accented Speech Recognition}
Existing studies on accented speech recognition mainly focus on applying acoustic features that are accent-invariant and an adaptation methods to allow the model to accommodate accented speech. \cite{zheng2005accent,najafian2014unsupervised} introduce acoustic features and adaptation method for recognizing accented speech. Meanwhile, \cite{jain2018improved,jain2019multi} and \cite{viglino2019end} explore a multi-task architecture that jointly learns an accent classifier and an acoustic model. \cite{jain2019multi} propose a mixture of expert models to segregate accent-specific and phone-specific speech variability in a joint framework, and \cite{sun2018domain} propose an adversarial training objective to help the model to learn accent-invariant features. In this work, we explore the possibility of recognizing speech with unseen accents, and extend MAML to enable fast adaptation by a few-shot learning in the cross-accent setting.

\section{Cross-Accented Speech Recognition}

In this section, we present the architecture of our transformer-based speech recognition model and the proposed meta-learning method for fast adaptation on the cross-accented speech recognition task.

\subsection{Transformer Speech Recognition Model}

\begin{figure}[!t]
  \centering
  \includegraphics[width=0.7\linewidth]{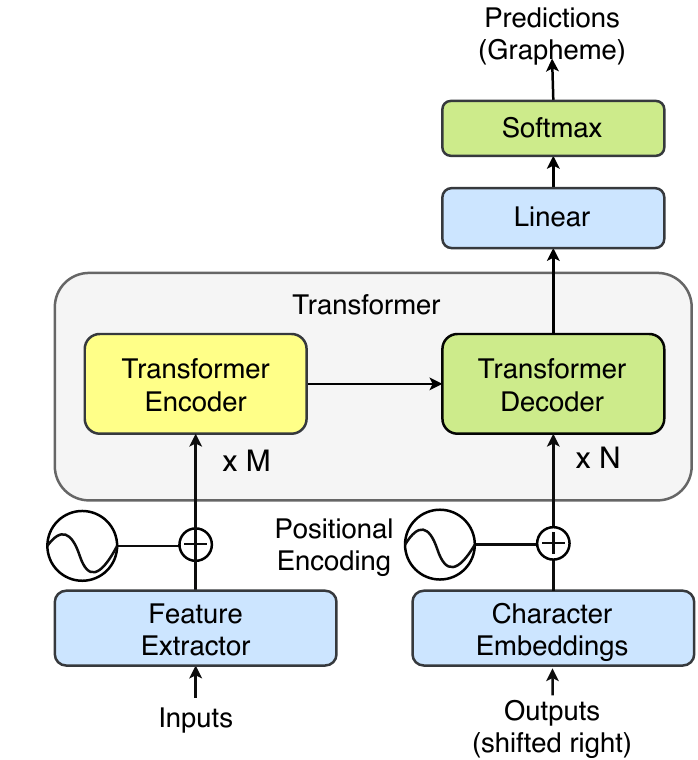}  
  \caption{Transformer ASR model architecture.}
  \label{fig:transformer-asr}
\end{figure}

We build our model using a sequence-to-sequence transformer ASR~\cite{vaswani2017attention,dong2018speech,winata2019code,winata2019lightweight} to learn to predict graphemes from the speech input. Our model extracts audio inputs with a learnable feature extractor module to generate input embeddings. The encoder uses input embeddings generated from the feature extractor module. Then the decoder receives the encoder outputs and applies multi-head attention to its input to finally calculates the logits of the outputs. To generate the probability of the outputs, we compute the softmax function of the logits. We apply a mask in the attention layer to avoid any information flow from future tokens, and we train our model by optimizing the next step prediction on the previous characters and by maximizing the log probability:
\begin{equation}
\max_{\theta}\sum_{i}\log{P(y_i|x,y^{\prime}_{<i};\theta)},
\end{equation}
\noindent where $x$ is the character inputs, $y_i$ is the next predicted character, and $y^{\prime}_{<i}$ is the ground truth of the previous characters. In the inference time, we generate the sequence using a beam-search in an auto-regressive manner. Then we maximize the following scoring function:
\begin{equation} \eta \sum_{i}\log{P(y_i|x,\hat{y}_{<i};\theta)} + \gamma \sqrt{wc(\hat{y}_{<i})},  
\end{equation}

\noindent where $\eta$ is the parameter to control the decoding probability from the decoder, and $\gamma$ is the parameter to control the effect of the word count $wc(\hat{y}_{<i})$ as suggested in~\cite{winata2019code} and \cite{winata2019lightweight}.

\begin{table}[!b]
\caption{Statistics of accented speech data in CommonVoice Dataset sorted alphabetically.}
\label{tab:statistics}
\centering
\resizebox{0.41\textwidth}{!}{
\begin{tabular}{l|cc}
\hline
\multirow{1}{*}{\textbf{accents}} & \multicolumn{1}{l}{\textbf{\# sample}} & \textbf{duration (hr)}  \\ \hline
Africa \textbf{(af)} & 4,065 & 5.04 \\
Australia \textbf{(au)} & 19,625 & 22.86 \\
Bermuda \textbf{(be)} & 363 & 0.46 \\
Canada \textbf{(ca)} & 17,422 & 20.20 \\
England \textbf{(en)} & 58,274 & 64.19 \\
Hong Kong \textbf{(hk)}  & 1,181 & 1.21 \\
India \textbf{(in)} & 23,878 & 29.09 \\
Ireland \textbf{(ir)} & 3,420 & 3.71 \\
Malaysia \textbf{(my)} & 843 & 1.07 \\
New Zealand \textbf{(nz)} & 6,070 & 7.06 \\
Philippines \textbf{(ph)} & 1,318 & 1.68 \\
Scotland \textbf{(sc)} & 4,376 & 5.08 \\
Singapore \textbf{(sg)} & 693 & 1.00 \\
South Atlantic \textbf{(sa)} & 212 & 0.23 \\
United States \textbf{(us)} & 145,692 & 163.89 \\ 
Wales \textbf{(wa)} & 1,128 & 1.16 \\ \hline
\textbf{Total} & 288,560 & 327.93 \\ \hline
\end{tabular}
}
\end{table}

\begin{table*}[!t]
\caption{Average Word Error Rate (\% WER) with Standard Error (SE) results in the mixed-region setting.}
\label{tab:wer_mixed_chart}
\centering
\resizebox{\textwidth}{!}{
\begin{tabular}{l|cccc|cccc}
\hline
\multirow{2}{*}{\textbf{accents}} & \multicolumn{4}{c|}{\textbf{MAML}} & \multicolumn{4}{c}{\textbf{Joint Training}}  \\ \cline{2-9} 
& $\tt{zero}$-$\tt{shot}$ & $\tt{5\%}$-$\tt{shot}$ & $\tt{25\%}$-$\tt{shot}$ &
$\tt{all}$-$\tt{shot}$ & $\tt{zero}$-$\tt{shot}$ & $\tt{5\%}$-$\tt{shot}$ & $\tt{25\%}$-$\tt{shot}$ & $\tt{all}$-$\tt{shot}$ \\ \hline
\multicolumn{9}{c}{\textit{without pre-training}} \\ \hline
Bermuda & 33.22 \footnotesize{$\pm$ 0.46} & 32.73 \footnotesize{$\pm$ 0.47} & 31.85 \footnotesize{$\pm$ 0.48} & \textbf{29.90 \footnotesize{$\pm$ 0.60}}
& 38.92 \footnotesize{$\pm$ 0.55} & 37.84 \footnotesize{$\pm$ 0.50} & 36.23 \footnotesize{$\pm$ 0.56} & 36.12 \footnotesize{$\pm$ 0.65} \\ \hline
Philippines & 50.08 \footnotesize{$\pm$ 0.56} & 48.22 \footnotesize{$\pm$ 0.69} & 45.94 \footnotesize{$\pm$ 0.64} & \textbf{44.43 \footnotesize{$\pm$ 0.69}} & 50.58 \footnotesize{$\pm$ 0.81} & 49.72 \footnotesize{$\pm$ 0.80} & 48.27 \footnotesize{$\pm$ 0.85} & 45.47 \footnotesize{$\pm$ 0.93} \\ \hline
Wales & 33.66 \footnotesize{$\pm$ 0.83} & 33.31 \footnotesize{$\pm$ 0.77} & 31.63 \footnotesize{$\pm$ 0.86} & \textbf{29.70 \footnotesize{$\pm$ 0.87}} & 37.04 \footnotesize{$\pm$ 0.68} & 37.43 \footnotesize{$\pm$ 0.69} & 35.60 \footnotesize{$\pm$ 0.80} & 32.37 \footnotesize{$\pm$ 0.87} \\ \hline
\multicolumn{9}{c}{\textit{with pre-training}} \\ \hline
Bermuda & 28.25 \footnotesize{$\pm$ 0.47} & 28.64 \footnotesize{$\pm$ 0.42} & 26.59 \footnotesize{$\pm$ 0.43} & \textbf{25.71 \footnotesize{$\pm$ 0.43}} & 31.42 \footnotesize{$\pm$ 0.57} & 31.43 \footnotesize{$\pm$ 0.56} & 30.05 \footnotesize{$\pm$ 0.44} & 27.64 \footnotesize{$\pm$ 0.40} \\ \hline
Philippines & 40.99 \footnotesize{$\pm$ 0.51} & 40.07 \footnotesize{$\pm$ 0.52} & 39.06 \footnotesize{$\pm$ 0.44} & \textbf{37.48 \footnotesize{$\pm$ 0.42}} & 43.17 \footnotesize{$\pm$ 0.83} & 41.98 \footnotesize{$\pm$ 0.76} & 40.56 \footnotesize{$\pm$ 0.77} & 38.79 \footnotesize{$\pm$ 0.69} \\ \hline
Wales & 25.91 \footnotesize{$\pm$ 0.73} & 25.55 \footnotesize{$\pm$ 0.86} & 23.94 \footnotesize{$\pm$ 0.73} & \textbf{23.40 \footnotesize{$\pm$ 0.64}} & 29.14 \footnotesize{$\pm$ 0.49} & 28.54 \footnotesize{$\pm$ 0.52} & 26.70 \footnotesize{$\pm$ 0.49} & 25.01 \footnotesize{$\pm$ 0.56} \\ \hline
\end{tabular}
}
\end{table*}

\subsection{Fast Adaptation via Meta-Learning}

Model-agnostic meta-learning (MAML)~\cite{finn2017model} learns to quickly adapt to a new task from a number of different tasks using a gradient descent procedure. In this paper, we apply MAML to effectively learn from a set of accents and quickly adapt to a new accent in the few-shot setting. We denote our Transformer ASR as $f_{\theta}$ parameterized by $\theta$. Our dataset is consist a set of accents $\mathcal{A} = \{A_1,A_2,\cdots,A_n \}$, and for each accent $i$, we split the data into $A_i^{tra}$ and $A_i^{val}$, then update $\theta$ into $
\theta^{\prime}$ by computing gradient descent updates on $A_i^{tra}$:
\begin{equation}
\theta_{i}^{\prime} = \theta - \alpha \nabla_{\theta}{\mathcal{L}_{A_{i}^{tra}}(f_{\theta})},
\end{equation}
\noindent where $\alpha$ is the fast adaptation learning rate. During the training, the model parameters are trained to optimize the performance of the adapted model $f(\theta_i^{\prime})$ on unseen $A_i^{val}$. The meta-objective is defined as follows:
\begin{align}
\min_{\theta}\sum_{A_i\sim p(\mathcal{A})}\mathcal{L}_{A_i^{val}}(f_{\theta_i'}) = \sum_{A_i\sim p(\mathcal{A})}\mathcal{L}_{A_i^{val}}(f_{\theta-\alpha \nabla_{\theta} \mathcal{L}_{A_i^{tra}}(f_{\theta})}).
\end{align}

\noindent where $\mathcal{L}_{A_i^{val}}(f_{\theta_i'})$ is the loss evaluated on $A_i^{val}$. We collect the loss $\mathcal{L}_{A_i^{val}}(f_{\theta_i'})$ from a batch of accents and perform the meta-optimization as follows:
\begin{align}
\label{eq:5}
\theta \leftarrow \theta - \beta {\sum_{A_i\sim p(\mathcal{A})}\nabla_{\theta}\mathcal{L}_{A_i^{val}}(f_{\theta_i'})},
\end{align}

\noindent where $\beta$ is the meta step size and $f_{\theta_i^{\prime}}$ is the adapted network on accent $A_i$. The meta-gradient update step is performed to achieve a good initialization for our model, then we can optimize our model with few number of samples on target accents in the fine-tuning step. In this work, we use first order approximation MAML as~\cite{gu2018meta} and~\cite{finn2018pmaml}, thus Equation \ref{eq:5} is reformulated as:
\begin{align}
\theta \leftarrow \theta - \beta {\sum_{A_i\sim p(\mathcal{A})}\nabla_{\theta_i'}\mathcal{L}_{A_i^{val}}(f_{\theta_i'})}.
\end{align}

\section{Experiments}

\subsection{Dataset}
We use the CommonVoice Dataset~\cite{ardila2019common},\footnote{We downloaded the data in December 2019} a multilingual open-accented dataset collected by Mozilla. In this work, we only use the English dataset and filter for only speech data with an accent label. There are 16 accents listed in the dataset, and we split the dataset into groups according to the accent label. The statistics of the English dataset are shown in Table~\ref{tab:statistics}. Note that the dataset is imbalanced and some accents only have very limited data. The pre-trained models are trained on the LibriSpeech corpus~\cite{panayotov2015librispeech}, a 960-hour training corpus of English read speech derived from audio books in the LibriVox project, sampled at 16 kHz. The accents are various and not labeled, but the majority are US English.\footnote{The LibriSpeech Dataset can be downloaded at http://www.openslr.org/12/ and the list of LibriVox accents can be found at https://wiki.librivox.org/index.php/Accents\_Table }

\subsection{Experimental Setup}
We preprocess raw audio input into a spectrogram before we fetch it into our model. Our model utilizes a VGG model~\cite{simonyan2014very}, a 6-layer CNN architecture, as the feature extractor. Our transformer model consists of two transformer encoder layers and four transformer decoder layers. The transformer consists of a $dim_{inner}$ of 2048, $dim_{model}$ of 512, and $dim_{emb}$ of 512. We use 8 heads for multi-head attention. In total, our model has around 10.2M parameters. For both the MAML and joint training models, we end the training process after 200k iterations. In the pre-training setting, we pre-train the model using the LibriSpeech Dataset for 1M iterations, and resume the training using the CommonVoice Dataset subsequently for another 100k iterations for all approaches. During the fine-tuning step, we run 10 iterations for each sample. We evaluate our model using a beam search with $\eta = 1$, $\gamma = 0.1$, and a beam size of 5. In the pre-training setting, we downsample the CommonVoice speech data to 16 kHz following the LibriSpeech Dataset audio sample rate. 

We train and evaluate the effectiveness of our fast adaptation method in two settings: (1) mixed-region setting, and (2) cross-region setting. The former is to train on ten accents, such as $\textbf{af}$, $\textbf{au}$, $\textbf{ca}$, $\textbf{en}$, $\textbf{hk}$, $\textbf{in}$, $\textbf{ir}$, $\textbf{my}$, $\textbf{nz}$, $\textbf{sa}$, $\textbf{sc}$, $\textbf{sg}$, and $\textbf{us}$, sampled from all regions, and we validate the model on the \textbf{ca}, \textbf{sc}, and \textbf{sa} accents and test the model on the \textbf{be}, \textbf{ph}, and \textbf{wa} accents. The latter is to train on five accents, such as \textbf{au}, \textbf{en}, \textbf{ir}, \textbf{nz}, and \textbf{us}, from specific regions and validate the model on the \textbf{ca}, \textbf{sc}, and \textbf{sa} accents, and test it on the \textbf{af}, \textbf{hk}, \textbf{in}, \textbf{ph}, and \textbf{sg} accents that come from other regions. We evaluate the model performance using the word error rate (WER) and run experiments ten times using different test folds. Each fold consists of 100 data randomly sampled from the test data. In the few-shot scenarios, we split the test accents data into training and testing sets. 75\% of the data are allocated for training, and the remainder for testing.\footnote{We will release the code and dataset manifests used in the experiments for reproducibility.} We report the average and standard error of all folds in zero-shot (0\%-shot), 5\%-shot, 25\%-shot, and all-shot (100\%-shot) settings. In addition, we also investigate the usefulness of pre-training on a large English corpus and fine-tune the model.

\begin{table*}[!t]
\caption{Average Word Error Rate (\% WER) with Standard Error (SE) results in the cross-region setting.}
\label{tab:wer_cross_chart}
\centering
\resizebox{\textwidth}{!}{
\begin{tabular}{l|cccc|cccc}
\hline
\multirow{2}{*}{\textbf{accents}} & \multicolumn{4}{c|}{\textbf{MAML}} & \multicolumn{4}{c}{\textbf{Joint Training}} \\ \cline{2-9} 
& $\tt{zero}$-$\tt{shot}$ & $\tt{5\%}$-$\tt{shot}$ & $\tt{25\%}$-$\tt{shot}$ &
$\tt{all}$-$\tt{shot}$ & $\tt{zero}$-$\tt{shot}$ & $\tt{5\%}$-$\tt{shot}$ & $\tt{25\%}$-$\tt{shot}$ & $\tt{all}$-$\tt{shot}$ \\ \hline
\multicolumn{9}{c}{\textit{without pre-training}} \\ \hline
Africa & 40.38 \footnotesize{$\pm$ 1.11} & 38.31 \footnotesize{$\pm$ 1.20} & 36.36 \footnotesize{$\pm$ 1.01} & \textbf{34.64 \footnotesize{$\pm$ 1.01}} & 41.56 \footnotesize{$\pm$ 1.04} & 41.40 \footnotesize{$\pm$ 1.08} & 39.34 \footnotesize{$\pm$ 1.34} & 38.32 \footnotesize{$\pm$ 1.17} \\ \hline
Hong Kong & 42.04 \footnotesize{$\pm$ 0.74} & 40.20 \footnotesize{$\pm$ 0.89} & 38.29 \footnotesize{$\pm$ 0.78} & \textbf{35.61 \footnotesize{$\pm$ 0.71}} & 44.84 \footnotesize{$\pm$ 0.65} & 44.88 \footnotesize{$\pm$ 0.67} & 44.09 \footnotesize{$\pm$ 0.66} & 41.28 \footnotesize{$\pm$ 0.59} \\ \hline
India & 62.07 \footnotesize{$\pm$ 0.90} & 54.60 \footnotesize{$\pm$ 1.46} & 51.71 \footnotesize{$\pm$ 1.06} & \textbf{47.85 \footnotesize{$\pm$ 1.00}} & 63.09 \footnotesize{$\pm$ 0.82} & 56.76 \footnotesize{$\pm$ 1.08} & 53.89 \footnotesize{$\pm$ 1.00} & 50.73 \footnotesize{$\pm$ 0.98} \\ \hline
Philippines & 50.06 \footnotesize{$\pm$ 0.74} & 48.17 \footnotesize{$\pm$ 0.71} & 47.71 \footnotesize{$\pm$ 0.78} & \textbf{45.05 \footnotesize{$\pm$ 0.82}} & 53.22 \footnotesize{$\pm$ 0.97} & 52.60 \footnotesize{$\pm$ 0.99} & 51.64 \footnotesize{$\pm$ 0.78} & 48.12 \footnotesize{$\pm$ 0.76} \\ \hline
Singapore & 55.75 \footnotesize{$\pm$ 0.85} & 55.76 \footnotesize{$\pm$ 0.83} & 54.43 \footnotesize{$\pm$ 0.68} & \textbf{52.71 \footnotesize{$\pm$ 1.06}} & 57.87 \footnotesize{$\pm$ 0.64} & 57.21 \footnotesize{$\pm$ 0.67} &  55.15 \footnotesize{$\pm$ 0.69} & 53.59 \footnotesize{$\pm$ 0.72}\\ \hline
\multicolumn{9}{c}{\textit{with pre-training}} \\ \hline
Africa & 32.63 \footnotesize{$\pm$ 1.25} & 31.75 \footnotesize{$\pm$ 1.19} & 31.09 \footnotesize{$\pm$ 1.22} & \textbf{29.75 \footnotesize{$\pm$ 1.01}} & 34.61 \footnotesize{$\pm$ 1.22} & 33.42 \footnotesize{$\pm$ 1.18} & 33.12 \footnotesize{$\pm$ 1.12} & 31.63 \footnotesize{$\pm$ 1.13} \\ \hline
Hong Kong & 36.06 \footnotesize{$\pm$ 0.56} & 36.04 \footnotesize{$\pm$ 0.71} & 32.38 \footnotesize{$\pm$ 0.71} & \textbf{32.15 \footnotesize{$\pm$ 0.62}} & 37.43 \footnotesize{$\pm$ 0.77} & 36.51 \footnotesize{$\pm$ 0.57} & 35.88 \footnotesize{$\pm$ 0.51} & 34.18 \footnotesize{$\pm$ 0.77}\\ \hline
India & 54.50 \footnotesize{$\pm$ 1.41} & 48.73 \footnotesize{$\pm$ 1.31} & 46.15 \footnotesize{$\pm$ 1.35} & \textbf{43.54 \footnotesize{$\pm$ 1.35}} & 55.43 \footnotesize{$\pm$ 1.36} &  50.52 \footnotesize{$\pm$ 1.26} & 48.63 \footnotesize{$\pm$ 1.32} & 46.58 \footnotesize{$\pm$ 1.07} \\ \hline
Philippines & 43.73 \footnotesize{$\pm$ 0.94} & 42.96 \footnotesize{$\pm$ 1.01} & 40.80 \footnotesize{$\pm$ 1.03} & \textbf{40.14 \footnotesize{$\pm$ 0.98}} & 45.16 \footnotesize{$\pm$ 0.98} & 44.64 \footnotesize{$\pm$ 1.04} & 42.38 \footnotesize{$\pm$ 0.88} & 41.74 \footnotesize{$\pm$ 0.98}\\ \hline
Singapore &  49.45 \footnotesize{$\pm$ 0.55} & 48.40 \footnotesize{$\pm$ 0.56} & 46.62 \footnotesize{$\pm$ 0.62} & \textbf{46.17 \footnotesize{$\pm$ 0.67}} & 52.06 \footnotesize{$\pm$ 0.71} & 50.48 \footnotesize{$\pm$ 0.70} & 49.43 \footnotesize{$\pm$ 0.69} & 47.11 \footnotesize{$\pm$ 0.66} \\ \hline
\end{tabular}
}
\end{table*}

\begin{figure}[!t]
  \centering
  \includegraphics[width=0.8\linewidth]{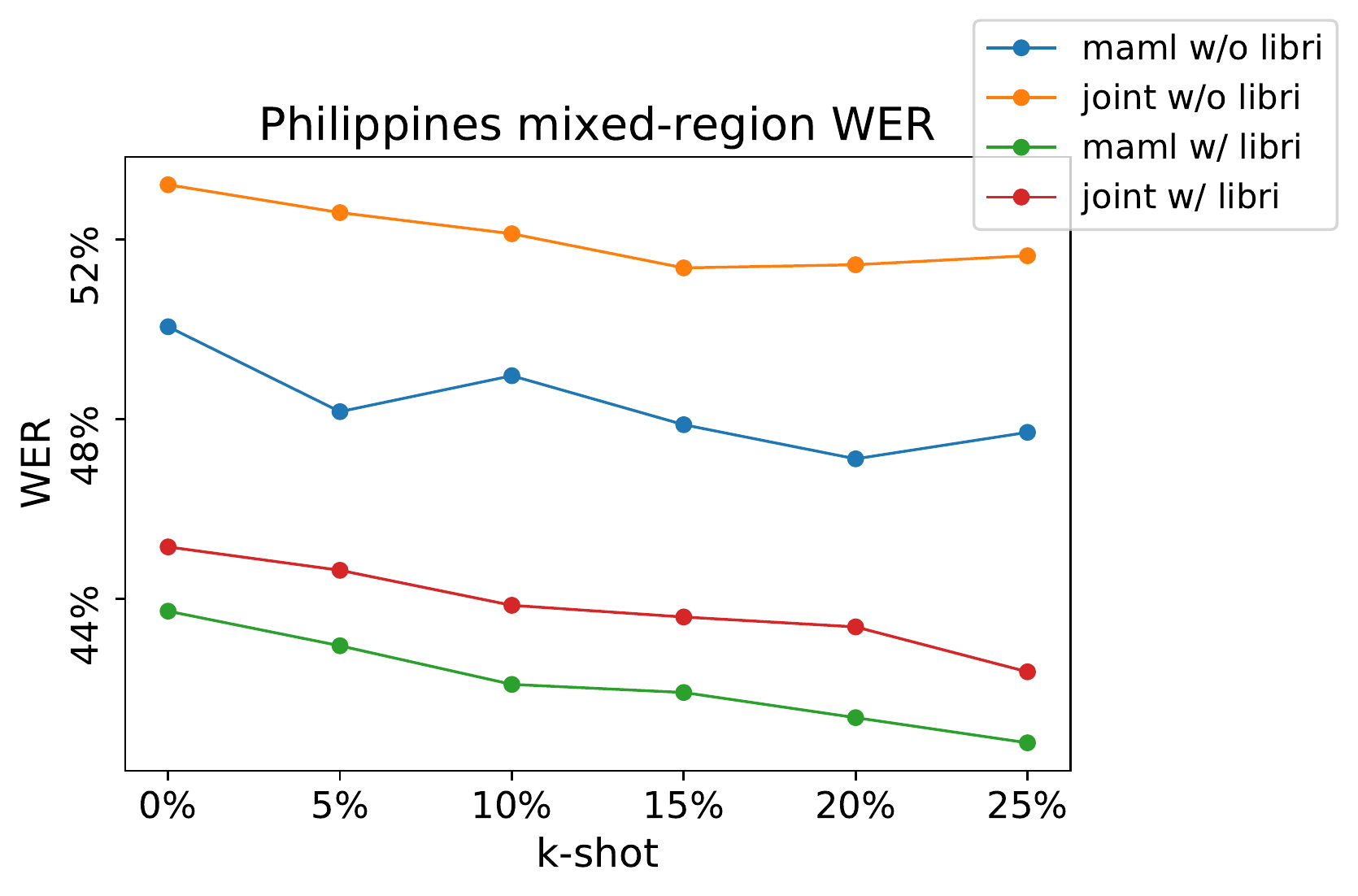}  
  \caption{Few-shot results on Philippines accent in the mixed-region setting.}
  \label{fig:wer_mixed_chart}
\end{figure}

\section{Results and Discussion}

\subsection{Quantitative Analysis}

As shown in Table~\ref{tab:wer_mixed_chart}, MAML consistently outperforms joint training in the mixed-region setting. The approach yields up to a 4\% WER margin in the zero-shot and few-shot settings. In general, for both MAML and joint-training, by adding more data on fine-tuning, the WER drops at a constant rate. Using the pre-trained model on the LibriSpeech Dataset significantly boosts the performance of all models by around 5\% to 8\% WER. In the all-shot setting, the results are similar to results in the 5\%-shot and 25\%-shot settings. We observe that the WER improvement after applying the pre-trained model for the Wales accent is higher than for the Bermuda and Philippines accents since the majority of the LibriSpeech Dataset is US accented speech which is far more acoustically similar to the accent of Wales than of Bermuda or Phillippines.


\subsection{Cross-region Performance}

We show the cross-region performance in Table~\ref{tab:wer_cross_chart}. As expected, the WER of the Philippines accent is slightly reduced when we remove Asian accents from the training data. Interestingly, focusing only on the Philippines accent results, as shown in Table~\ref{tab:wer_mixed_chart} and Table~\ref{tab:wer_cross_chart}, MAML on the cross-region setting yielding WER performance similar to the joint-training on the mixed-region setting. The same result is not shown from training on the cross-region setting. Based on the empirical results, we can conclude that MAML is far more accent-agnostic compared to joint training. In sum, the model trained with MAML performs better than with joint training and learns more accent-invariant representations.

\begin{figure}[!h]
  \centering
  \includegraphics[width=0.8\linewidth]{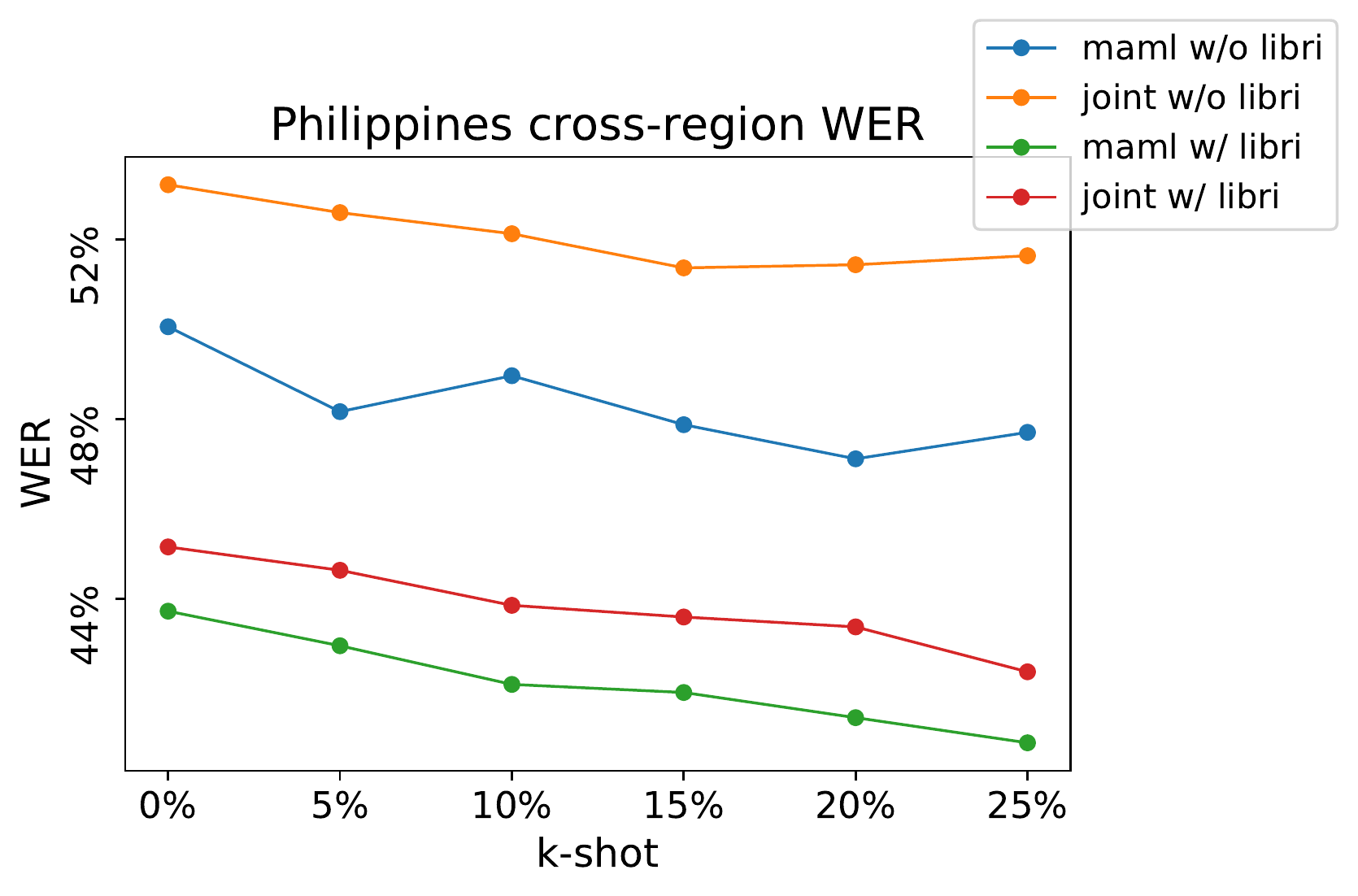} 
  \caption{Few-shot results on Philippines accent in the cross-region setting.}
  \label{fig:wer_cross_chart}
\end{figure}

\subsection{Effectiveness of Few-Shot Fine-tuning}
We first investigate the number of samples needed to start showing performance improvement after fine-tuning. We start by training the model with a very small number of samples, from one to ten, where each sample approximately consists of 4 seconds of audio. We observe that the model does not adapt to the target accent with a miniscule amount of data. We believe that our model is not able to capture the information from a very short audio sample due to a large acoustic variation in the data. Therefore, we increase the minimum threshold to 5\% of the training data, and the model starts to adapt to the target accent accordingly.

In Figure \ref{fig:wer_mixed_chart} and Figure \ref{fig:wer_cross_chart}, in general, MAML performs better than joint training in all settings. By having more target accented speech data, the model gains higher performance with a lower WER for both the mixed-accent and cross-accent settings. We observe that MAML is effectively applied to models without pre-training on the LibriSpeech Dataset and it decays much faster than joint training. 

We further investigate the effectiveness of fast-adaptability of the MAML approach compared to the all-shot setting. As shown in Tables \ref{tab:wer_mixed_chart} and  \ref{tab:wer_cross_chart}, the MAML approach with 25\%-shot fine-tuning performs similarly or even better compared to the joint approach with all-shot fine tuning, both in the mixed-accent and cross-accent settings. In the all-shot setting, the MAML approach can further improve the performance, and outperforms the joint training approach in all experiment settings. In light of the impressive experimental results of the MAML approach, we can infer that MAML has fast adaptability to low-resource unseen accented data.




\section{Conclusions}

In this paper, we introduce a cross-accented speech recognition task derived from an existing dataset, CommonVoice, and establish a new benchmark for evaluating cross-accented speech recognition in the mixed-region and cross-region scenarios.
We apply a fast adaptation method via the model-agnostic meta-learning (MAML) approach to learn a robust speech recognition system to rapidly adapt to unseen accents. Based on the empirical results, MAML consistently outperforms the non-meta learning baseline in all settings around 4\% WER improvement compared to joint training in both the mixed-region and cross-region scenarios. 
Impressively, MAML leverages less data (25\%-shot) and achieves comparable results to joint-training with all training data (all-shot).
We also further improve the performance of our model by adding pre-training on a large speech corpus.



\section{Acknowledgements}

This work has been partially funded by ITF/319/16FP and MRP/055/18 of the Innovation Technology Commission, the Hong Kong SAR Government, and School of Engineering Ph.D. Fellowship Award, the Hong Kong University of Science and Technology, and RDC 1718050-0 of EMOS.AI.

\bibliographystyle{IEEEtran}

\bibliography{mybib}

\begin{thebibliography}{10}
\providecommand{\url}[1]{#1}
\csname url@samestyle\endcsname
\providecommand{\newblock}{\relax}
\providecommand{\bibinfo}[2]{#2}
\providecommand{\BIBentrySTDinterwordspacing}{\spaceskip=0pt\relax}
\providecommand{\BIBentryALTinterwordstretchfactor}{4}
\providecommand{\BIBentryALTinterwordspacing}{\spaceskip=\fontdimen2\font plus
\BIBentryALTinterwordstretchfactor\fontdimen3\font minus
  \fontdimen4\font\relax}
\providecommand{\BIBforeignlanguage}[2]{{%
\expandafter\ifx\csname l@#1\endcsname\relax
\typeout{** WARNING: IEEEtran.bst: No hyphenation pattern has been}%
\typeout{** loaded for the language `#1'. Using the pattern for}%
\typeout{** the default language instead.}%
\else
\language=\csname l@#1\endcsname
\fi
#2}}
\providecommand{\BIBdecl}{\relax}
\BIBdecl

\bibitem{kat1999fast}
L.~W. Kat and P.~Fung, ``Fast accent identification and accented speech
  recognition,'' in \emph{1999 IEEE International Conference on Acoustics,
  Speech, and Signal Processing. Proceedings. ICASSP99 (Cat. No. 99CH36258)},
  vol.~1.\hskip 1em plus 0.5em minus 0.4em\relax IEEE, 1999, pp. 221--224.

\bibitem{narayanan2014joint}
A.~Narayanan and D.~Wang, ``Joint noise adaptive training for robust automatic
  speech recognition,'' in \emph{2014 IEEE International Conference on
  Acoustics, Speech and Signal Processing (ICASSP)}.\hskip 1em plus 0.5em minus
  0.4em\relax IEEE, 2014, pp. 2504--2508.

\bibitem{hori2017advances}
T.~Hori, S.~Watanabe, Y.~Zhang, and W.~Chan, ``Advances in joint ctc-attention
  based end-to-end speech recognition with a deep cnn encoder and rnn-lm,''
  \emph{Proc. Interspeech 2017}, pp. 949--953, 2017.

\bibitem{finn2017model}
C.~Finn, P.~Abbeel, and S.~Levine, ``Model-agnostic meta-learning for fast
  adaptation of deep networks,'' in \emph{Proceedings of the 34th International
  Conference on Machine Learning-Volume 70}.\hskip 1em plus 0.5em minus
  0.4em\relax JMLR. org, 2017, pp. 1126--1135.

\bibitem{sun2018domain}
S.~Sun, C.-F. Yeh, M.-Y. Hwang, M.~Ostendorf, and L.~Xie, ``Domain adversarial
  training for accented speech recognition,'' in \emph{2018 IEEE International
  Conference on Acoustics, Speech and Signal Processing (ICASSP)}.\hskip 1em
  plus 0.5em minus 0.4em\relax IEEE, 2018, pp. 4854--4858.

\bibitem{jain2018improved}
A.~Jain, M.~Upreti, and P.~Jyothi, ``Improved accented speech recognition using
  accent embeddings and multi-task learning.'' in \emph{Interspeech}, 2018, pp.
  2454--2458.

\bibitem{jain2019multi}
A.~Jain, V.~P. Singh, and S.~P. Rath, ``A multi-accent acoustic model using
  mixture of experts for speech recognition,'' \emph{Proc. Interspeech 2019},
  pp. 779--783, 2019.

\bibitem{ardila2019common}
R.~Ardila, M.~Branson, K.~Davis, M.~Henretty, M.~Kohler, J.~Meyer, R.~Morais,
  L.~Saunders, F.~M. Tyers, and G.~Weber, ``Common voice: A
  massively-multilingual speech corpus,'' \emph{arXiv preprint
  arXiv:1912.06670}, 2019.

\bibitem{vaswani2017attention}
A.~Vaswani, N.~Shazeer, N.~Parmar, J.~Uszkoreit, L.~Jones, A.~N. Gomez,
  {\L}.~Kaiser, and I.~Polosukhin, ``Attention is all you need,'' in
  \emph{Advances in neural information processing systems}, 2017, pp.
  5998--6008.

\bibitem{gu2018meta}
J.~Gu, Y.~Wang, Y.~Chen, V.~O. Li, and K.~Cho, ``Meta-learning for low-resource
  neural machine translation,'' in \emph{Proceedings of the 2018 Conference on
  Empirical Methods in Natural Language Processing}, 2018, pp. 3622--3631.

\bibitem{schmidhuber1992learning}
J.~Schmidhuber, ``Learning to control fast-weight memories: An alternative to
  dynamic recurrent networks,'' \emph{Neural Computation}, vol.~4, no.~1, pp.
  131--139, 1992.

\bibitem{thrun2012learning}
S.~Thrun and L.~Pratt, \emph{Learning to learn}.\hskip 1em plus 0.5em minus
  0.4em\relax Springer Science \& Business Media, 2012.

\bibitem{ravi2016optimization}
S.~Ravi and H.~Larochelle, ``Optimization as a model for few-shot learning,''
  \emph{ICLR}, 2016.

\bibitem{vinyals2016matching}
O.~Vinyals, C.~Blundell, T.~Lillicrap, D.~Wierstra \emph{et~al.}, ``Matching
  networks for one shot learning,'' in \emph{Advances in neural information
  processing systems}, 2016, pp. 3630--3638.

\bibitem{santoro2016meta}
A.~Santoro, S.~Bartunov, M.~Botvinick, D.~Wierstra, and T.~Lillicrap,
  ``Meta-learning with memory-augmented neural networks,'' in
  \emph{International conference on machine learning}, 2016, pp. 1842--1850.

\bibitem{yu2018diverse}
M.~Yu, X.~Guo, J.~Yi, S.~Chang, S.~Potdar, Y.~Cheng, G.~Tesauro, H.~Wang, and
  B.~Zhou, ``Diverse few-shot text classification with multiple metrics,'' in
  \emph{Proceedings of the 2018 Conference of the North American Chapter of the
  Association for Computational Linguistics: Human Language Technologies,
  Volume 1 (Long Papers)}, 2018, pp. 1206--1215.

\bibitem{madotto2019paml}
\BIBentryALTinterwordspacing
A.~Madotto, Z.~Lin, C.-S. Wu, and P.~Fung, ``Personalizing dialogue agents via
  meta-learning,'' in \emph{Proceedings of the 57th Annual Meeting of the
  Association for Computational Linguistics}.\hskip 1em plus 0.5em minus
  0.4em\relax Florence, Italy: Association for Computational Linguistics, Jul.
  2019, pp. 5454--5459. [Online]. Available:
  \url{https://www.aclweb.org/anthology/P19-1542}
\BIBentrySTDinterwordspacing

\bibitem{qian-yu-2019-domain}
\BIBentryALTinterwordspacing
K.~Qian and Z.~Yu, ``Domain adaptive dialog generation via meta learning,'' in
  \emph{Proceedings of the 57th Annual Meeting of the Association for
  Computational Linguistics}.\hskip 1em plus 0.5em minus 0.4em\relax Florence,
  Italy: Association for Computational Linguistics, Jul. 2019, pp. 2639--2649.
  [Online]. Available: \url{https://www.aclweb.org/anthology/P19-1253}
\BIBentrySTDinterwordspacing

\bibitem{huang2018natural}
P.-S. Huang, C.~Wang, R.~Singh, W.-t. Yih, and X.~He, ``Natural language to
  structured query generation via meta-learning,'' in \emph{Proceedings of the
  2018 Conference of the North American Chapter of the Association for
  Computational Linguistics: Human Language Technologies, Volume 2 (Short
  Papers)}, 2018, pp. 732--738.

\bibitem{lin2019learning}
Z.~Lin, A.~Madotto, G.~I. Winata, Z.~Liu, Y.~Xu, C.~Gao, and P.~Fung,
  ``Learning to learn sales prediction with social media sentiment,'' in
  \emph{Proceedings of the First Workshop on Financial Technology and Natural
  Language Processing}, 2019, pp. 47--53.

\bibitem{hsu2019meta}
J.-Y. Hsu, Y.-J. Chen, and H.-y. Lee, ``Meta learning for end-to-end
  low-resource speech recognition,'' \emph{arXiv preprint arXiv:1910.12094},
  2019.

\bibitem{klejch2018learning}
O.~Klejch, J.~Fainberg, and P.~Bell, ``Learning to adapt: A meta-learning
  approach for speaker adaptation,'' \emph{Proc. Interspeech 2018}, pp.
  867--871, 2018.

\bibitem{klejch2019speaker}
O.~Klejch, J.~Fainberg, P.~Bell, and S.~Renals, ``Speaker adaptive training
  using model agnostic meta-learning,'' \emph{arXiv preprint arXiv:1910.10605},
  2019.

\bibitem{zheng2005accent}
Y.~Zheng, R.~Sproat, L.~Gu, I.~Shafran, H.~Zhou, Y.~Su, D.~Jurafsky, R.~Starr,
  and S.-Y. Yoon, ``Accent detection and speech recognition for
  shanghai-accented mandarin,'' in \emph{Ninth European Conference on Speech
  Communication and Technology}, 2005.

\bibitem{najafian2014unsupervised}
M.~Najafian, A.~DeMarco, S.~Cox, and M.~Russell, ``Unsupervised model selection
  for recognition of regional accented speech,'' in \emph{Fifteenth annual
  conference of the international speech communication association}, 2014.

\bibitem{viglino2019end}
T.~Viglino, P.~Motlicek, and M.~Cernak, ``End-to-end accented speech
  recognition,'' \emph{Proc. Interspeech 2019}, pp. 2140--2144, 2019.

\bibitem{dong2018speech}
L.~Dong, S.~Xu, and B.~Xu, ``Speech-transformer: a no-recurrence
  sequence-to-sequence model for speech recognition,'' in \emph{2018 IEEE
  International Conference on Acoustics, Speech and Signal Processing
  (ICASSP)}.\hskip 1em plus 0.5em minus 0.4em\relax IEEE, 2018, pp. 5884--5888.

\bibitem{winata2019code}
G.~I. Winata, A.~Madotto, C.-S. Wu, and P.~Fung, ``Code-switched language
  models using neural based synthetic data from parallel sentences,'' in
  \emph{Proceedings of the 23rd Conference on Computational Natural Language
  Learning (CoNLL)}, 2019, pp. 271--280.

\bibitem{winata2019lightweight}
G.~I. Winata, S.~Cahyawijaya, Z.~Lin, Z.~Liu, and P.~Fung, ``Lightweight and
  efficient end-to-end speech recognition using low-rank transformer,''
  \emph{arXiv preprint arXiv:1910.13923}, 2019.

\bibitem{finn2018pmaml}
\BIBentryALTinterwordspacing
C.~Finn, K.~Xu, and S.~Levine, ``Probabilistic model-agnostic meta-learning,''
  in \emph{Proceedings of the 32Nd International Conference on Neural
  Information Processing Systems}, ser. NIPS'18.\hskip 1em plus 0.5em minus
  0.4em\relax USA: Curran Associates Inc., 2018, pp. 9537--9548. [Online].
  Available: \url{http://dl.acm.org/citation.cfm?id=3327546.3327622}
\BIBentrySTDinterwordspacing

\bibitem{panayotov2015librispeech}
V.~Panayotov, G.~Chen, D.~Povey, and S.~Khudanpur, ``Librispeech: an asr corpus
  based on public domain audio books,'' in \emph{2015 IEEE International
  Conference on Acoustics, Speech and Signal Processing (ICASSP)}.\hskip 1em
  plus 0.5em minus 0.4em\relax IEEE, 2015, pp. 5206--5210.

\bibitem{simonyan2014very}
K.~Simonyan and A.~Zisserman, ``Very deep convolutional networks for
  large-scale image recognition,'' in \emph{ICLR}, 2015.

\end{thebibliography}


\end{document}